\documentclass[twocolumn,showpacs,preprintnumbers,amsmath,amssymb]{revtex4}
\usepackage{graphicx}
\usepackage{graphicx}
\usepackage{dcolumn}
\usepackage{bm}
\begin{document}
\title
{
Possible origin of shoulder in the reactor antineutrino spectrum
}
\author{Ajit Kumar Mohanty}
\email{ajitkumar.mohanty@saha.ac.in}

\affiliation{Saha Institute of Nuclear Physics, Kolkata - 700064, Bhabha Atomic Research Centre, Mumbai-400084 and HBNI, Mumbai}
\date{\today}

\begin{abstract}
{
The re-analysis of measured $^{238}U$ integrated electron spectrum 
reveals a  shape mismatch with respect to the so called {\it ab initio}
calculation and the ratio resembles well with the recently observed 
 reactor antineutrino shoulder or 
bump around 5 MeV prompt energy when plotted in a reduced scale
where the first two moments (area and average) of the spectrum are  made identical (KNO type scaling). However, such bumps are not seen in
case of other actinides like $^{235}U$ and $^{239,241}Pu$ isotopes. 
Since the magnitude of the bump in $^{238}U$ is comparable with the experimental observations, $^{238}U$ can not be
the only source of anomaly unless similar contributions come from other uranium and plutonium isotopes as well. Therefore, it is
strongly conjectured that the integrated electron spectra of uranium and plutonium isotopes particularly in the 
$6$ MeV to $8$ MeV energy range  may have additional contributions coming from the
harder part of the reactor neutron spectrum which may not be present in the thermal measurements. 

}
\end{abstract}

\maketitle

All modern reactor antineutrino flux measurements are associated with two important puzzles: one related to the
deficit of the total antineutrino yields of the short base line measurements and the other
one to the  spectral shape that deviates significantly at around $6$ to $8$ MeV antineutrino energy in comparison to the
standard predictions (commonly referred to as Huber-Muller model) \cite{muller, huber}.
Quite often  the first puzzle is interpreted with the  
possible existence of a fourth
neutrino which is sterile in nature in addition to the three standard model light neutrinos that participate
in the oscillations \cite{mention}. If sterile neutrino exists, the expected range of mass and mixing angle 
have been constrained severely by many  
recent measurements and analysis  \cite{aartsen,adamson1,adamson2,collin}.
The Daya Bay measurements of antineutrino flux with changing fuel burn-ups \cite{an1}
provide additional challenge to the
sterile neutrino hypotheses by showing that
the deficit has isotopic dependence, although the interpretation is still not
free from ambiguities  \cite{hayes3,giunti_ji,gebre}.
The second puzzle not necessarily related to the former  was first reported by RENO collaboration \cite{seo,choi}
and subsequently reconfirmed by Daya Bay \cite{an}, Double Chooz \cite{abe}  and NEOS \cite{ko} collaborations
respectively. In a nuclear reactor, the antineutrinos are produced by successive beta decays of neutron rich
fragments coming predominantly from the fission of four ($^{235}U$, $^{238}U$, $^{239}Pu$ and $^{241}Pu$)
isotopes. The above experiments
detect these antineutrinos directly through the inverse beta decay process (IBD: $\bar \nu  + p \rightarrow e^++n$) 
by measuring the  
emitted prompt gamma energy due to the positron annihilation followed by thermal neutron capture.
It has been reported by the above collaborations that the ratio of the measured spectrum with respect
to the Huber-Muller model prediction shows a distinct peak at about $5$ MeV prompt energy. This is the second
puzzle as mentioned above since the event normalized ratio is expected to be unity at all energies.
It may be pointed out here that unlike direct antineutrino 
measurements using IBD, all model calculations estimate the antineutrino flux indirectly 
either by using a conversion  method
or an {\it ab initio} method 
by explaining the measured integrated beta spectrum. The integrated beta spectrum for 
$^{235}U$, $^{239}Pu$ and $^{241}Pu$ were measured in mid 80's using high flux thermal 
reactor at ILL(Institut Laue-Langevin), Grenoble \cite{u235,pu239,pu241} and more recently (in 2014) for $^{238}U$ using
fast neutron beam \cite{u238}. The corresponding 
antineutrino spectra have been derived from these measurements by using a conversion
method elaborated more detail in \cite{u235,pu239,pu241} while in \cite{u238}, the antineutrino spectrum conversion 
for $^{238}U$ 
has been carried
out using an empirical
relation between $N_{\bar \nu}$ (number of antineutrinos per fission per MeV)  and $N_\beta$ (number of electrons
per fission per MeV) given by,

\begin{center}
\begin{eqnarray}
\label{eq1}
N_{\bar \nu}(E)=k(E)N_\beta(E-\delta),
\end{eqnarray}
\end{center}

where $\delta$ is close to  $m_e c^2$ and the function $k(E)$ is expected to be unity.
On the other hand, the {\it ab initio}  method commonly known as summation method requires complete knowledge
of fission fragment yields of each isotope and their subsequent decay channels until each fission product reaches 
equilibrium. The beta spectrum (and corresponding antineutrino spectrum) is estimated as the sum over all beta 
branches and subsequently as the sum over all fission fragments with proper weight factors. Therefore, the
summation method heavily depends on the information available in the nuclear data bases.  Although, the summation
method is more elegant, it suffers from large uncertainty either due to incomplete information or erroneous measurements
of nuclear data. One of the common source of uncertainty is the incorrect assignment of the strength of the beta branching ratios 
of the nucleus which are prone to the pandemonium effects unless measurements are carried out using total absorption
spectroscopy (TAS) \cite{fallot,zakari,rasco}. There could also be uncertainty in the fission yield measurements \cite{wilson} 
as well as due to lack of knowledge of several forbidden transitions \cite{hayes1,sonzogni_Mc}. In 2011, Muller et al reported
the new predictions for antineutrino flux for the above four isotopes based on an  
{\it ab initio} calculation by using an improved data base (ENSDF plus pandemonium corrected nuclei) which
agrees with the ILL electron data within a few percent accuracy. The remaining discrepancies are corrected by fitting the ILL data
with a set of five virtual beta branches \cite{muller}. These results are reconfirmed later on by Huber \cite{huber}. The
Huber-Muller estimations have remained so far as the most upto date and state of the art 
predictions and is being used as the reference point
by all experiments which have reported anomaly \cite{an,abe,ko}.      
Several attempts have been made in the recent years to understand this anomaly with different interpretations 
\cite{dwyer,hayes2,sonzogni1,sonzogni2}. Some argue that $^{238}U$ is responsible for  the bump \cite{hayes2} which
is ruled out for the fact that this isotope contributes only about less than $\sim 10\%$ to the total nuclear power production in 
a reactor and alone can not reproduce the magnitude of the bump. On the other hand, a recent analysis of Huber using NEOS and
Daya Bay data favors $^{235}U$ as the isotope contributing to the bump \cite{huber1} which is corroborated by the
fact that the IBD yield of  $^{235}U$ deviates the most from the predictions \cite{giunti,an1}. 

Although, future measurements may shed some light
on the importance of the individual isotopes  \cite{buck}, in the present work, we reanalyze the    
most recently measured $^{238}U$ electron and derived antineutrino spectra \cite{u238}.  
A shape mismatch in the form of a bump is noticed just by comparing the experimental data with the so called {\it ab initio}
calculation.  This bump resembles well with the recently observed 5 MeV reactor antineutrino bump when plotted using
KNO type reduced scale
where the first two moments (area and average) of the spectra are  made identical . However, such bumps are not found in 
other actinides. 
Since the magnitude of the bump in $^{238}U$ is comparable with the experimental observations, $^{238}U$ can not be
the only source of anomaly unless similar contributions come from other uranium and plutonium isotopes also. Therefore, it is
conjectured that the integrated electron spectra of uranium and plutonium isotopes measured at the reactor particularly in the
$6$ MeV to $8$ MeV energy range  may have additional contributions coming from the
harder part of the neutron spectrum which may not be present in the thermal measurements.

\begin{figure}
    \includegraphics[width=\linewidth]{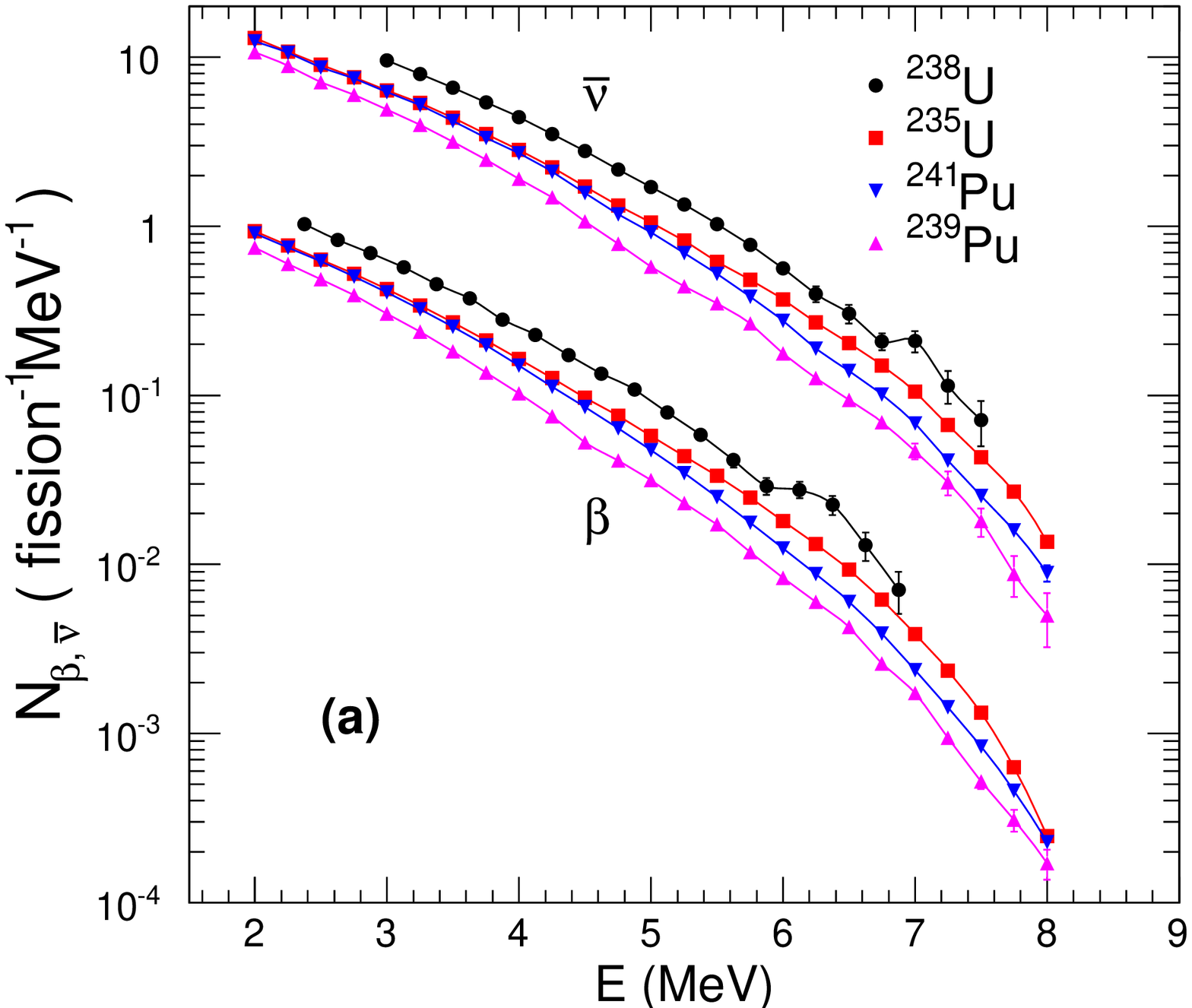}
    \includegraphics[width=\linewidth]{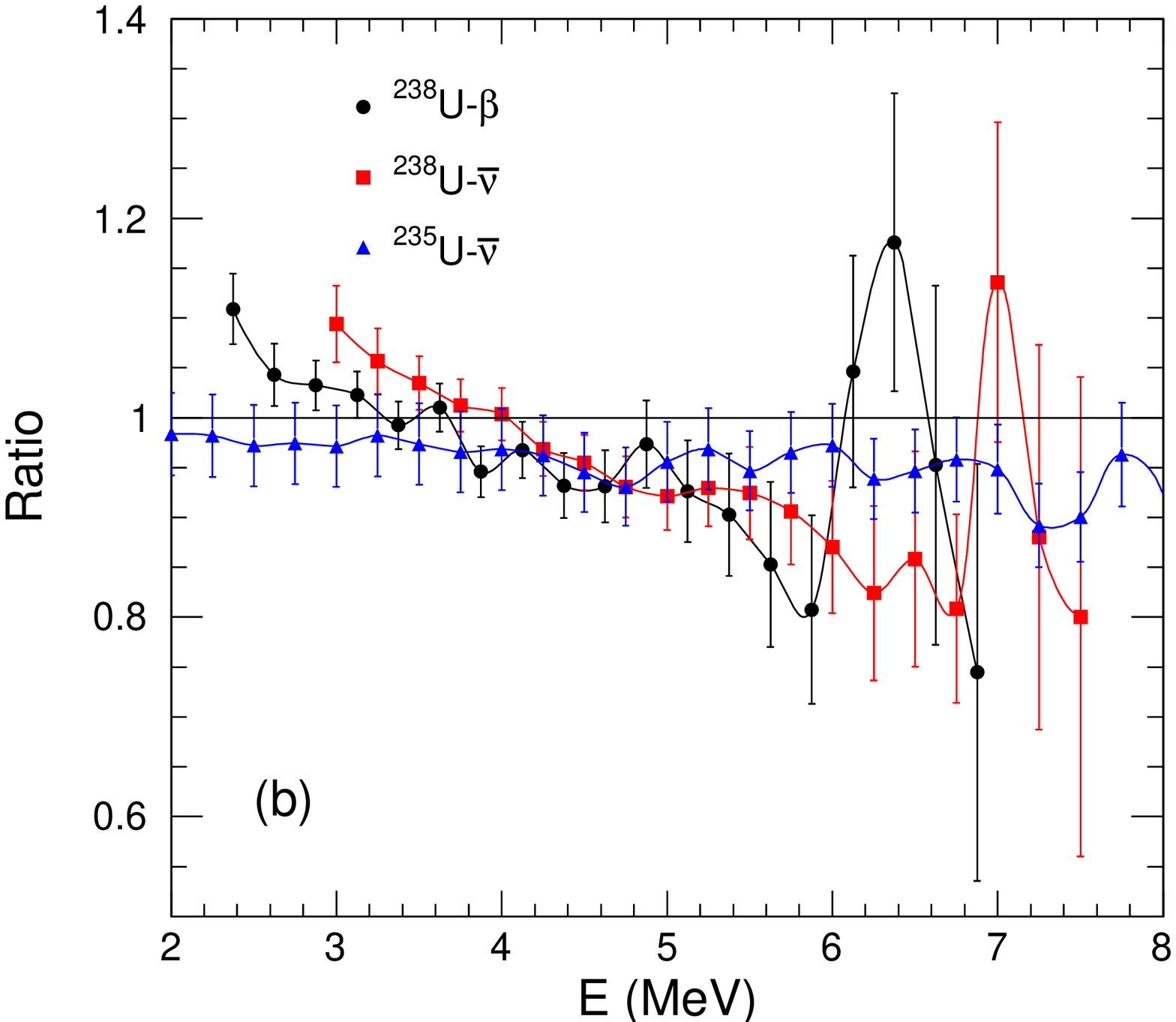}
    \caption{\label{fig1}
(a) The experimental  $N_{\beta, \bar \nu}$ (number of electron or antineutrino 
per fission per MeV) as a function of
kinetic energy.  The data points for  $^{235}U$, $^{241}Pu$ and $^{238}U$  are taken from
Refs. \cite{u235},\cite{pu241} and \cite{u238} respectively. For $^{239}Pu$, the data points are taken from
Ref. \cite{pu239} and Ref. \cite{pu241} respectively. The $N_{\bar \nu}$ spectra are scaled up by a factor
of $10$ for clarity. The plots are restricted to the energy range $2$ MeV to $8$ MeV only.
The Figure (b) shows the ratio with
respect to the Muller-Huber predictions as described in the text. 
The curves join the data points smoothly to guide the eye. 
}
\end{figure}

Before proceeding further, we first re-examine the experimental data points once again. For comparison, Figure \ref{fig1}(a) shows
$N_{\beta,\bar \nu}$ (number of electron or antineutrino per fission per MeV) 
as a function of their kinetic energies for four isotopes. The solid curve joins two successive data points smoothly just to guide the eye.
Although the measurements for $^{238}U$ have large error bars, a simple 
visual observation shows the existence of a  shoulder like structure at around $6$ MeV to $8$ MeV
electron or antineutrino energies whereas such bumpy structures are not seen in case of other three isotopes. Next, 
we estimate the ratio of the measurements with respect to
the corresponding Huber-Muller predictions obtained using the following parametrization,

\begin{eqnarray}
\label{eq2}
N(E)=exp \big ( \sum_{n=0}^6 a_n E^{n-1} \big ),
\end{eqnarray}
where the parameters $a_n$ for antineutrions are taken directly from \cite{muller} and \cite{huber}. However, there is no listed 
parameters for predicted electron spectra except for $^{238}U$ which has been tabulated in \cite{muller}.    
Therefore, we fit this predicted beta spectrum using the above parametrization assuming an energy 
independent error of
$3\%$. The fit parameters of beta spectrum for $^{238}U$ are listed in the second column of table I.    
The third and fourth columns list the parameters of antineutrino for $^{238}U$ and $^{235}U$ taken from \cite{muller}
and \cite{huber} respectively for easy reference. 
Figure \ref{fig1}(b) shows the ratio plot for $^{238}U$ both for electron and antineutrino and ratio plot of 
$^{235}U$ only for antineutrino
as a function of kinetic energy. The behavior of plutonium isotopes are similar to $^{235}U$ and not shown in the figure.
As can be seen from the figure \ref{fig1}(b), both the ratios 
of electron and antineutrino for $^{238}U$ show a peak structure at around $6$ MeV 
to $8$ MeV where as such peak is not seen in case of $^{235}U$.

\begin{table}[ht]
\caption{The parameters $a_0$ to $a_5$  
as defined in the text are listed below. The first column lists the parameters for experimental $\beta$
spectrum of $^{238}U$. The data points are taken from \cite{u238}.
The second column lists the $\beta$ parameters for $^{238}U$ obtained fitting the predicted data points as listed
in third column of table III in Ref. \cite{muller} with inclusion of nominal $3\%$ error.
The third and fourth columns list the parameters for $\bar \nu$ for $^{238}U$ and $^{235}U$ taken from \cite{muller}
and \cite{huber} respectively. 
}
\begin{center}
\begin{tabular}{|c|c|c|c|}
\hline
U238-Expt & U238-Muller & U238-Muller & U235-Huber\\   
$\beta$ & $\beta$ & $\bar \nu$ & $\bar \nu$\\   
\hline
6.588 & 0.4619 & 4.833(-1) & 4.367 \\
\hline
-7.253 & 0.184 & 1.927(-1) & -4.577 \\
\hline
3.416 & -0.1776 & -1.283(-1) & 2.100 \\
\hline
-0.8586 &-3.97(-2) & -6.762(-3)& -5.294(-1) \\
\hline
0.1023 & 3.255(-3) & 2.233(-3) & 6.186(-2) \\
\hline
-4.719(-3) & -2.585(-4) & -1.536(-4) & -2.777(-3) \\
\hline
\end{tabular}
\end{center}
\end{table}

Being
encouraged by this simple observation in case of $^{238}U$, we now proceed to see how well this peak compares with the
experimental measurements.
To make the comparison more sensitive, we use a reduced scale where the x-axis is scaled down and y-axis is scaled up by 
the average $<E>$. More explicitly, $E$ is replaced by  $E/<E>$ and $N_{\beta, \bar \nu}$ by,

\begin{eqnarray}
N_{\beta, \bar \nu}~\big (E \big) \rightarrow \frac{<E>}{A}~N_{\beta, \bar \nu}~\big (\frac{E}{<E>}\big),
\end{eqnarray}
where $A$ is the area under the curve. 
This is similar to the KNO scaling used in high energy physics to study particle productions when it was
first observed by Koba-Nielsen-Olesen (KNO) that particle productions in certain type of process exhibit an universal
scaling when ploted in a reduced scale \cite{kno}. 
Without going into the physics aspect of the KNO
scaling, it can be said that on the KNO scale,  all the plots have unity area and unity average value. In other words, the first
two moments of the curves (area and average) under KNO scaling are made identical so that any shape differences if any
can be seen more conveniently in the reduced scale.

\begin{figure}
    \includegraphics[width=\linewidth]{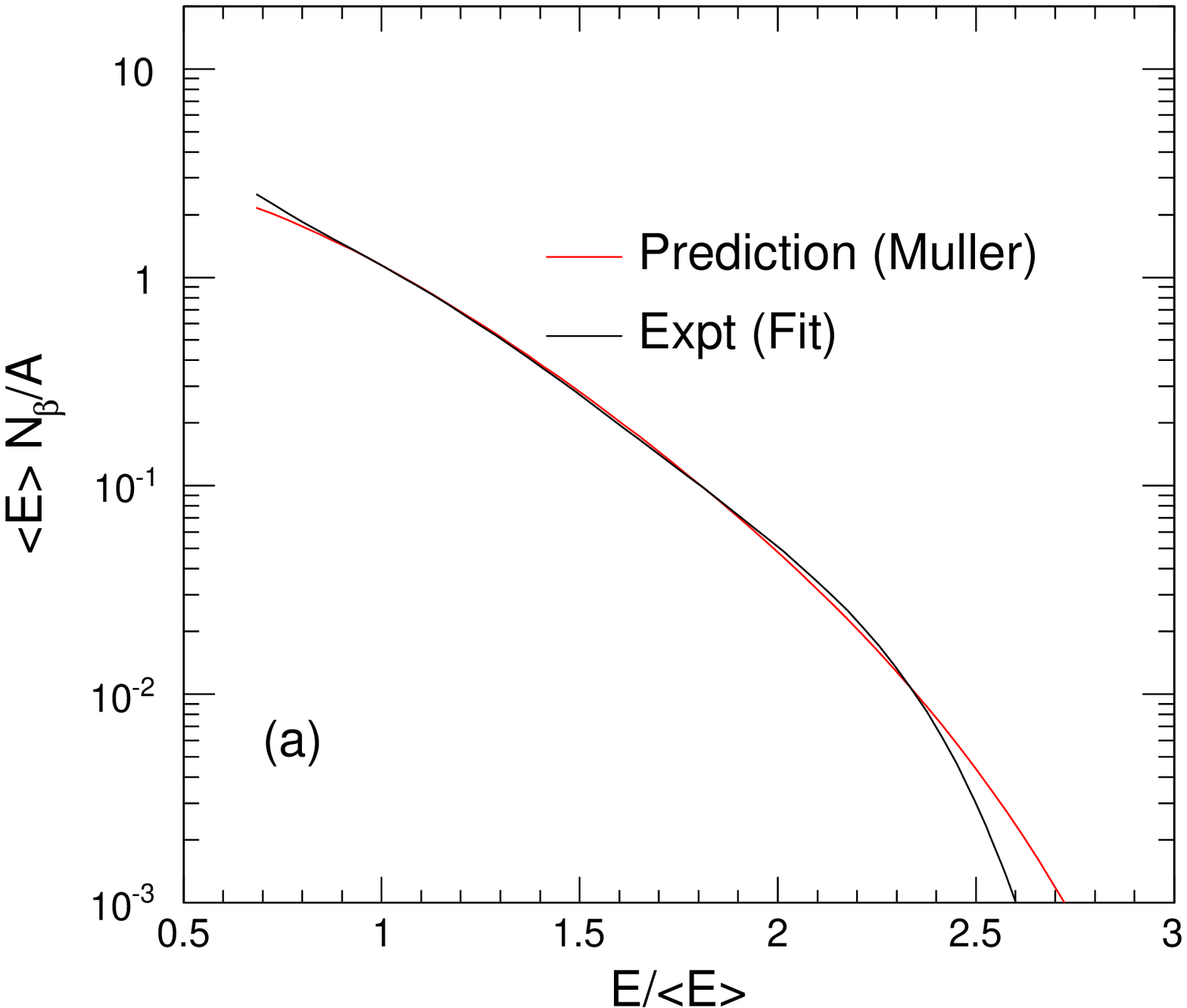}
    \includegraphics[width=\linewidth]{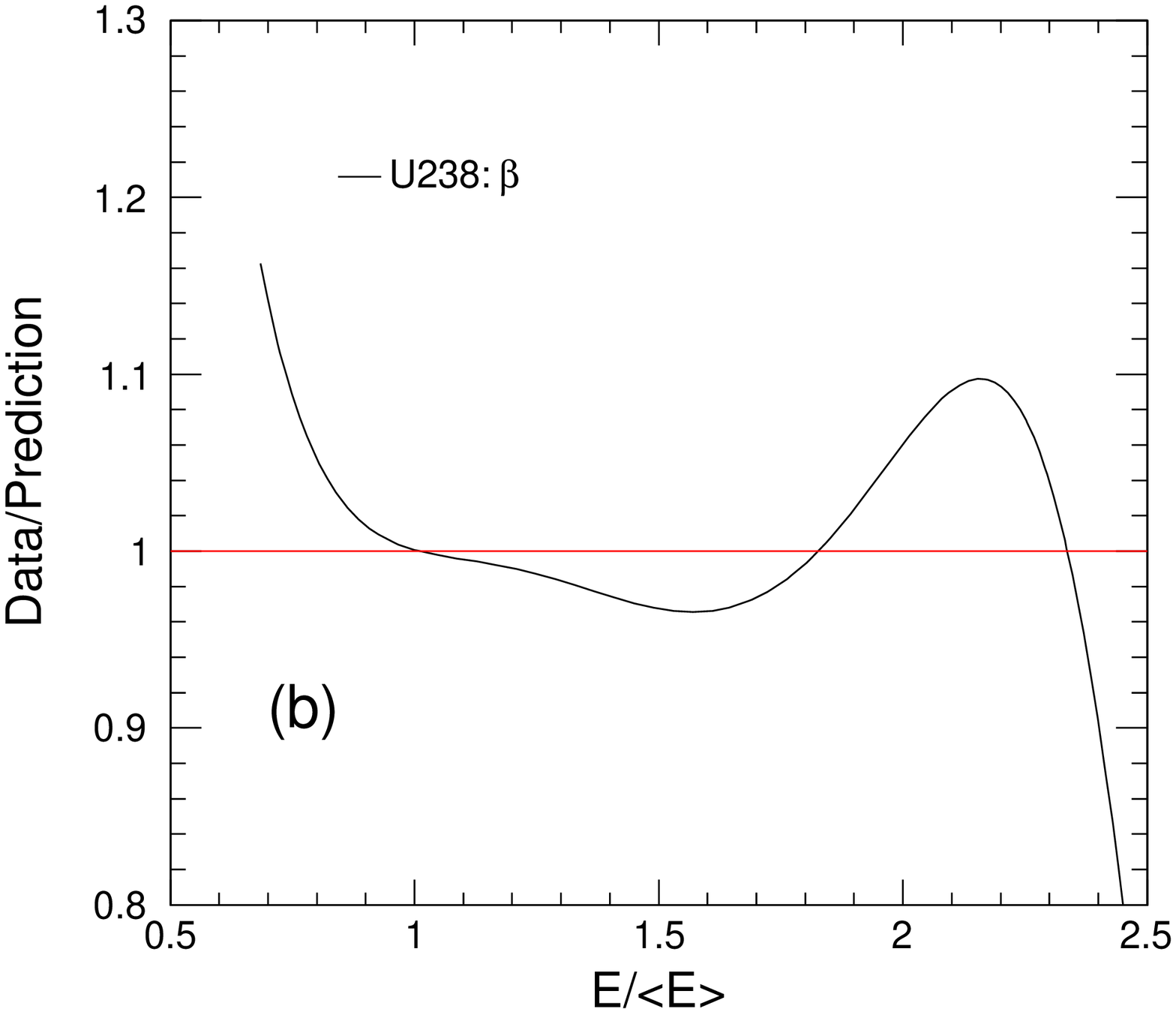}
    \caption{\label{fig2}
The  top panel (a) shows the $\beta$ spectrum  on a dimensionless reduced scale as described in the text.
The black curve is obtained using the parameters as given in the first column of the table I
while the red curve is obtained using the parameters as given in second column corresponding to Muller et al
prediction \cite{muller}. 
The bottom panel Figure 2(b)
shows the ratio of the fitted and predicted values. 
}
\end{figure}

Instead of using discrete points, we fit the $^{238}U$ experimental electron data points \cite{u238} using above parametrization with
inclusion of statistical errors only. We get a reasonably good fit with $\chi^2/ndf \sim 10.5/13$ and the 
parameters are listed in the first column of table I. For comparison
with the predictions, we restrict the analysis in the range $2$ MeV to $8$ MeV only. In this range, the fitted experimental curve
has an area $A_1 \sim 1.61$ (number of electrons per fission)  with an average energy $<E_1> \sim 3.03$ MeV where as the corresponding 
predicted values are $A_2 \sim 1.54$ (number of electrons per fission) 
 and $<E_2> \sim 3.1$ MeV  respectively. Using these values, both the curves are rescaled and  plotted
in Figure \ref{fig2}. Although in the reduced scale, both the curves have unity area and unity average values, they have considerable
shape differences which is evident in the ratio plot as shown in the bottom panel (b). This ratio can be compared with the corresponding
ratio in Figure \ref{fig1} (see the black curve) for $^{238}U$ beta spectrum. In Figure \ref{fig1}, the peak appears at around
$E \sim 6.5$ MeV which corresponds to $E/<E>$ $\sim 2.2$ in the reduced scale as expected. It is also seen that after the peak, the
ratio falls of sharply with increasing energy as the predicted curve falls of less steeply as compared to the fitted curve. While more
precise measurements are needed at higher energies, it may be pointed out here that data points for electron spectrum is available only from
$2.375$ MeV to $6.875$ MeV (taking the mid point of the energy bin) and the extrapolation may not remain valid at very high energy.

\begin{figure}
    \includegraphics[width=\linewidth]{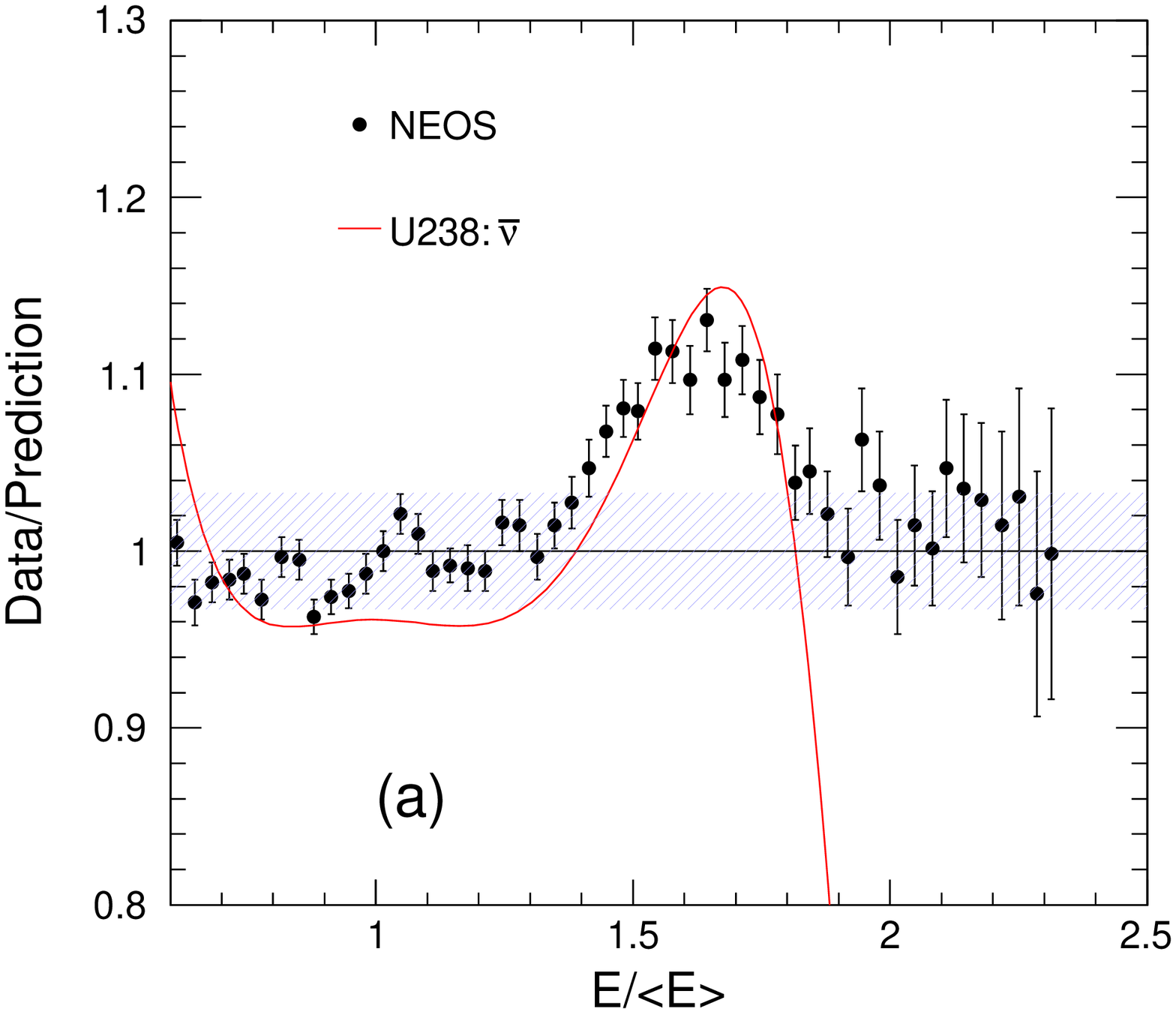}
    \includegraphics[width=\linewidth]{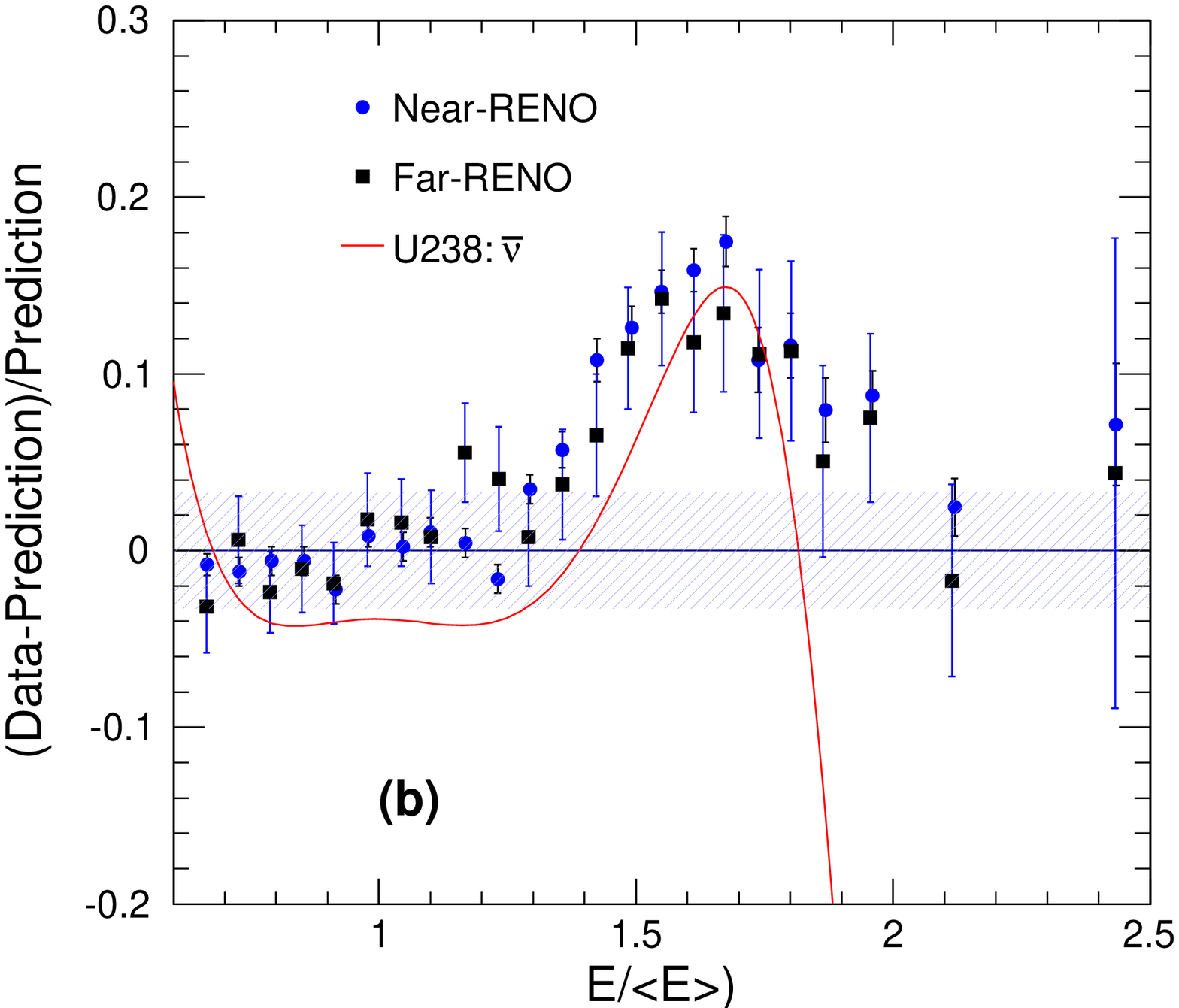}
    \caption{\label{fig3}
      The ratio versus kinetic energy of antineutrino obtained 
using the procedure as discussed in the text. The data points are
taken from NEOS (top pannel)  \cite{ko} and RENO experiments (bottom panel) \cite{choi} respectively. 
The hatched area corresponds to $3.3\%$ absolute normalization error.
}
  \end{figure}

Next, we estimate the antineutrino spectrum from the experimental electron spectrum using the same conversion
procedure as given in \cite{u238} (Eq.\ref{eq1} above) 
and compare the ratios with the reported experimental measurements. More explicitly, we calculate 
the quantity $\sigma(E)~N_{\bar \nu}(E)$ where $\sigma$ is the antineutrino interaction cross section given by,
(in the leading order) \cite{vogel}

\begin{eqnarray}
\sigma = 0.0952 \times 10^{-42}~cm^2~(E_ep_e/~1MeV^2).
\end{eqnarray}

where $E_e = E_{\bar \nu} -(M_n-M_p)$ is the positron energy when neutron recoil
energy is ignored and $p_e$ is the positron momentum. 
As in \cite{u238}, we use
$\delta=0.561$ MeV and evaluate $k(E)$ using Muller et al predictions \cite{muller}. Finally, the ratios are estimated using the reduced
scale as shown in Figure \ref{fig3}. The top panel shows the ratios for NEOS collaboration \cite{ko} and the bottom
panel for the RENO collaboration \cite{choi} (both near and far detectors) respectively. The experimental prompt energy 
spectrum has a mean value of  about $<E> \sim 3.0$  MeV for NEOS measurements and $<E> \sim 3.2$ MeV for RENO measurements. 
For comparison, the x-axis of the data points are scaled down
by the corresponding $<E>$.
As can be seen, the ratio resembles well with the 
measurements except at higher energies where the discrepancy arises due to  
either the experimental measurements falls off sharply  or the Huber-Muller
prediction over estimates the values. While the above discrepancies at higher energy require further investigation with more precision
measurements, it is interesting to note that the strength of the peak obtained from $^{238}U$ alone is comparable to the experimental
measurements although $^{238}U$ contributes not more than $\sim 10\%$ to the total reactor fission power generation. It is expected that
if the peak is due to $^{238}U$ alone, this height (data/prediction) should have been at least $\sim 2$ \cite{huber1}. 
Therefore, it is unlikely that $^{238}U$ alone
can contribute to the desired strength unless nearly equal contributions come from other uranium and plutonium isotopes as well.
It may be mentioned here that the hint for a bump in the measured $^{238}U$ spectrum at more or less the right energy was noticed,
in \cite{vogel1}. However, we give a more quatitative estimate of the excess in a model independent way which compares
well with the reported experimental ratios.

In conclusions, we have noticed a bumpy structure in $^{238}U$ both in the measured
beta spectrum as well as in the derived antineutrino spectrum
when compared with the Muller et al (more precisely Huber-Muller)  predictions. Subsequently, 
for analysis purpose, we represent the experimental data points by a
suitable parametrization which is obtained by fitting the
experimental electron spectrum with inclusion of statistical errors as quoted in \cite{u238}. For a better sensitivity and
also to predict the maximum possible deviation, we rescaled the spectrum using  a KNO type dimensionless scaling where first two moments
of the curve are made unity. In the reduced scale, 
it is noticed that the estimated ratio of $^{238}U$ compares well with the reported measurements. 
However, such deviations
are not noticed in case of other fissioning isotopes. Since $^{238}U$ 
contributes about $\sim 10\%$ to the total fission power generation in a nuclear
reactor, the ratio (data/prediction) should have been $\sim 2$ had it been the only isotope responsible for the observed anomaly. 
Therefore, it is highly conjectured that
the other isotopes particularly $^{235}U$ and $^{239}Pu$ must be contributing significantly to this anomaly. 
Since the antineutrino spectra obtained from the
measured electron spectra for these isotopes do not show any anomaly when compared with Huber-Muller predictions, it       
is possible that the integrated electron spectra of uranium and plutonium isotopes particularly in the 
$6$ MeV to $8$ MeV energy range  may have additional contributions coming from the
harder part of the neutron spectrum which may not be present in the thermal measurements. 
Therefore,
it is necessary to measure the integrated beta spectrum for all isotopes undergoing neutron induced fission 
covering energies from thermal to epithermal regions.

\section*{Acknowledgments}
\addcontentsline{toc}{section}{Acknowledgements}
{We would like to thank Dr. Dipak Mishra  for many fruitful  discussions and critical comments.}

\end{document}